\documentclass{article}
\usepackage{graphics}
\begin{document}

\newcommand{\bc}{\begin{center}}
\newcommand{\ec}{\end{center}}
\newcommand{\be}{\begin{equation}}
\newcommand{\ee}{\end{equation}}
\newcommand{\no}{\noindent}
\newcommand{\vs}{\vspace{2.5mm}}
\newcommand{\vn}{\vs\no}
\newcommand{\rmd}{{\rm d}}
\newcommand{\rme}{{\rm e}}
\newcommand{\rarr}{\rightarrow}
\newcommand{\x}{\mbox{$\times$}}

 \baselineskip 6 mm
 \bc
 {\LARGE\bf Can we observe galaxies that recede faster than
light ? --- A more clear-cut answer}

\vs\vs{\large  T. Kiang}

{\textit{Dunsink Observatory, Dublin Institute for Advanced
Studies,\\ Dublin 15, Ireland}}\ec 

\baselineskip 4.46 mm
\begin{quotation}

\vs\vs\no{\bf Abstract}~~A more clear-cut answer to the
title question is:  Yes, if the expanding universe started
with a big bang; No, if it started infinitely slowly.  

\vs\vs\no{\bf Key words:}~~cosmology---cosmological models 
\end{quotation}

\vn{\bf 1. Formulations of 
the Question}~~The present Letter is a
follow-up of a previous, much longer paper (Kiang 1997) on
the same question stated in the title.  The verbs in the
question were left deliberately vague as to tense, because
the previous paper dealt with both of the two formulations, 
(i) ``Can we {\it now} observe galaxies that {\it were
receding\/} faster than light at the time of emission~?'' and
(ii) ``Can we {\it eventually\/} observe galaxies that are
{\it now} receding  faster than light~?''

In the present Letter I shall concentrate on the first
formulation, which seems to be of more practical interest,
for it can be re-stated in a sharper form: ``Among the objects
we now observe can we identify those that were receding
faster than light at the time of emission ?''

It appears that, when so formulated, the question is capabale
of quite a clear-cut answer. That is the main thrust of this
Letter. But before I present it, I shall (1) summarize the
relevant results so far obtained and (2) consider the
question for the currently favored ($\Omega_M=0.3,\,
\Omega_{\Lambda}=0.7,\, \Omega_K=0$) model (the ``30/70 model''
hereinafter). It was the result obtained with the 30/70 model
that prompted me to seek, and eventually to find, the more
general result.

\vn{\bf 2. Previous Results}

The main relevant results of my (1997) paper are as follows.

1. This question of observability has nothing to do with
{\em horizon}. 

2. It depends solely on the form of the scale factor, $R(t)$,
of the cosmological model. More precisely, it simply depends
on whether or not two well-defined curves in the $r-t$ plane
($r$ the coordinate distance, $t$ the cosmic time) intersect
at some point in the past. The two curves are, 1) the curve
that separates the regions of subluminal and superluminal
recession velocities,
\be r_{v=c}(t) = 1 / \dot{R}(t) \;,  \ee

\no and 2) the equation for our past light cone:
\be r_{\rm PLC}(t)=\int_t^{t_0} \frac{\rmd t}{R(t)}\;, \ee

\no $t_0$ being the present epoch. Here and throughout this Letter, 
we always use $c=1$ units.  

3. It then followed that for the steady-state model, the
answer is ``no''; on the other hand, for all the three
varieties ($k=0, \pm 1$) of the big bang ($\Lambda=0$) model,
the answer is ``yes''; in particular, for the $k=0$ model
(the "standard" big bang), we have the result that all
quasars with redshifts greater than 1.25 (and we now know
thousands of such objects) had superluminal recession
velocities at the time of emission.

\vn {\bf 3. The 30/70 Model}~~The equations appear simplest
when we use the Hubble units (Kiang 1987), i.e., when all times
are in units of the current Hubble time $H_0^{-1}$ and all
distances, the current Hubble radius $cH_0^{-1}$. Then, $c$
and $H_0$ will not appear in the equations (being equal
to 1), and all the physical quantities are non-dimensional.
In these units, the Friedmann equation for the 30/70 model
is, (cf.\ e.g., Davis et al., 2003, Equation (11); Bondi
1960, p.80),  
\be  \dot{R}^2 = 0.3\,R^{-1} + 0.7\,R^2 \;, \ee

\no  and we have the explicit solution (Bondi 1960, p.82,
putting $C=0.3,\, \lambda=3\x 0.7=2.1$ in the Bondi formula 
for his case 2 (i)),   
\be  R^3(t') = 0.2143\,[\cosh(2.510\,t')-1] \;, \ee

\no where $t'$ is time reckoned from the big bang.  Since for
$t'=0.964$ we have $R=1$, the age of the universe in this 
model is $t_0=0.964\,H_0^{-1}$. In the figures that follow, I
shall be using time reckoned from the present epoch, $t$,
rather than $t'$; we have $t'=t+0.964$.

From these expressions it is easy to calculate the two curves
(1) and (2). These are shown in Fig.\,1, along with the scale
factor $R$. The scale factor for the standard big bang
model is included for interest;---it is labelled $R_1$:

\begin{figure}
\centering {\includegraphics{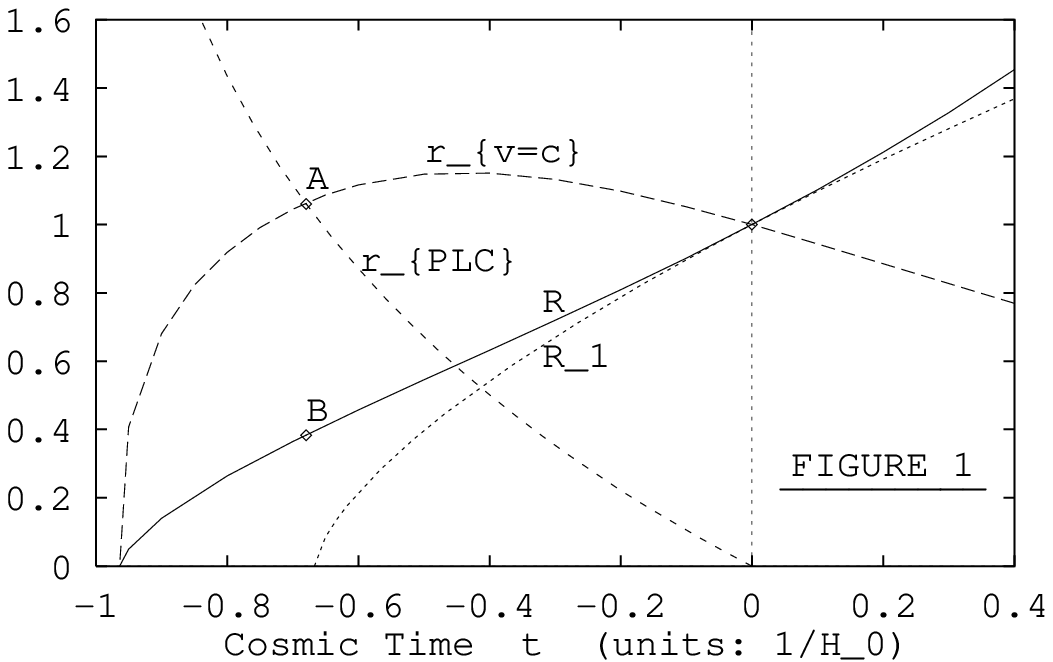}}
\bc{\small Fig.1~~The scale factor $R(t)$, and the $r_{v=c}(t)$
and $r_{\rm PLC} (t)$ curves for the 30/70 model.    $R_1(t)$
is the scale factor for the standard big bang model}   
\ec
\end{figure}

\vs The $v=c$ locus (1) and our past light cone (2) intersect
at point A $(t=-0.681,\, r=1.063)$. At $t=-0.681$, we have
$R=0.383$ (Point B). This last value corresponds to $z=1.61$.
Hence we may assert that 
according to the 30/70 model, all objects with redshifts
greater than 1.61 have the property that, at the time of
emission, they were receding from us faster than light. 

This result is qualitatively the same as for the standard
big bang model; the only difference is that, in the
latter, the dividing redshift is even lower, at 1.25.

\vn{\bf 4. Towards a More Clear-Cut Answer}. It thus appears
that the answer to the stated question 
may not depend on 
whether or not there is a non-zero cosmological constant
$\Lambda$, that the answer 
is always
``yes'', as long as the
universal expansion started with a big bang, that is, as long
as there was a point in the finite past, $t_{\rm BB}$, when
$R=0$ and $\dot{R}=+\infty$. This 
conjecture is easily proved to be true as follows.
Consider the behavior of the two curves (1) and (2) as
we move backward in time from $t=0$ to $t=t_{\rm BB}$.
Consider curve (2) first. Since $R(t)$ is always positive,
$r_{\rm PLC}$ must start from 0 and increase
monotonically to some finite value 
at $t_{\rm BB}$ 
(The integral in (2) converges at $t_{\rm BB}$, because 
generally, $R \rarr (t')^{2/3}$ as $t'\rarr 0$).
On the other hand, the $r_{v=c}$ curve (1) always starts at
unity (when the Hubble units are used) at $t=0$, and ends at
0 at $t=t_{\rm BB}$, whether the decrease is monotonic, as in
$\Lambda=0$ cases, or whether it first increases then
decreases, as may happen in a $\Lambda\neq 0$ case. In all
cases, the two curves must intersect at some time between the
present epoch and $t_{\rm BB}$, and we have proved

\begin{quotation}
\vn{\bf Lemma 1}~~{\em If the universal expansion started with
a big bang, then of the objects we now observe, those with
redshifts above a certain calculable value have the property
that, at emission, they were receding from us at speeds
greater than $c$.}
\end{quotation}

\vs Now, what happens if the universe did not start with a bang,
but very, very gently from time minus infinity$\:$?  The
steady-state model is one such model and I found in my 1997
paper that here, the $r_{v=c}$ curve lies consistently one
Hubble radius above the $r_{\rm PLC}$ curve; in fact, for
all $t<0$, 
\be r_{v=c}(t)=1+\rme^{|t|},\; 
          r_{\rm PLC}(t)=\rme^{|t|}\;. \ee

\no Hence, as already stated, the answer is ``no'' here.

Now, the steady-state model is a non-relativistic model. Of
the relativistic models (more precisely, Friedmann models),
only those with a non-negative mass parameter ($\Omega_M \geq
0$) and a non-negative curvature parameter ($\Omega_K \geq 0$)
and the corresponding critical dark-energy parameter
($\Omega_{\Lambda}=
  \Omega_{\Lambda_c}=(4\,\Omega_K^3) / (27\,\Omega_M^2)$) start
expansion infinitely slowly  from
a  value of $R$ equal to $(3\,\Omega_M) / (2\,\Omega_K)$ (Bondi
p.82). 
Felten and Isaacman (1986) 
have comprehensively reviewed all Friedmann models with
non-negative mass density $\Omega_M$ 
for all values of
the cosmological constant $\Lambda$.

Fig.\,2 shows the $R(t)$ for the model ($\Omega_M=0.1,
\Omega_K=0.45, \Omega_{\Lambda}=1.35$), as a typical
representative of $\Lambda=\Lambda_c$ models.  Also shown is
the scale factor, $R^{\ast}(t)$, for the limiting case 
($\Omega_M=\Omega_K=0, \Omega_{\Lambda_c}=1$).

\begin{figure}
\centering {\includegraphics{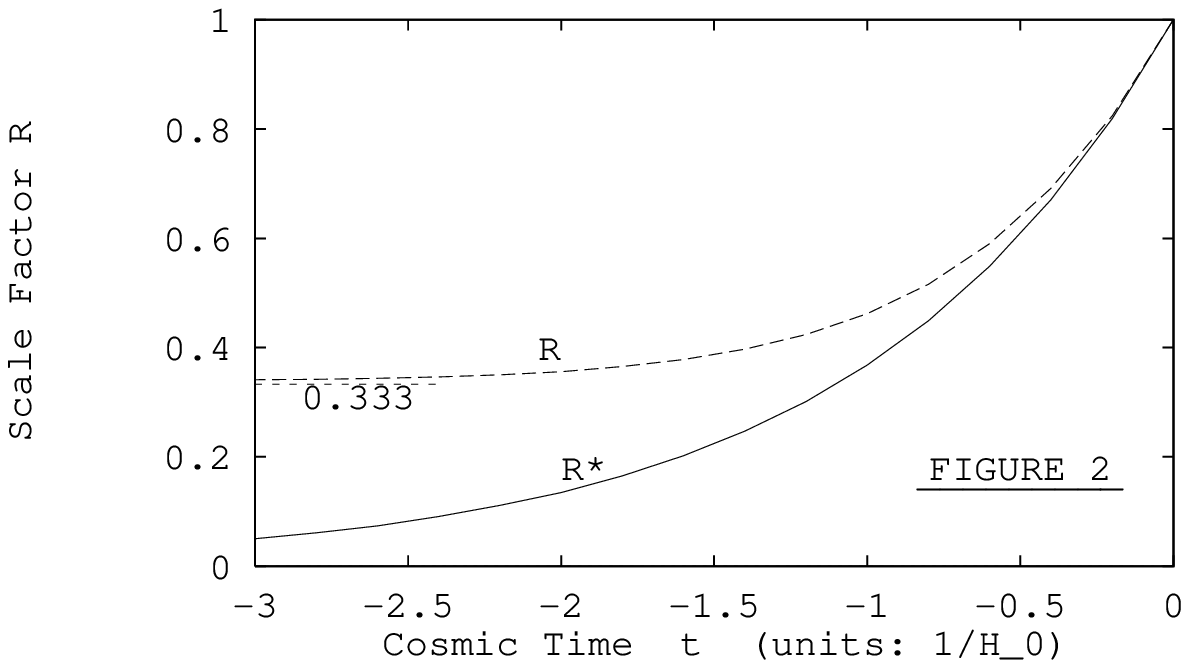}}
\bc
{\small Fig.2~~$R(t)$ is the scale factor for the model
$(\Omega_M=0.1, \Omega_K=0.45, \Omega_{\Lambda_c}=1.35)$,
$R^{\ast}(t)$ is that for the limiting model 
$(\Omega_M = \Omega_K = 0, \Omega_{\Lambda_c}=1)$}
\ec
\end{figure}

The limiting (0,0,1) model is just the de-Sitter model. Its scale
factor is the same as that of the steady-state model: $R^{\ast}(t)
= \exp (t)$ for all $t$ between $-\infty$ and $+\infty$.
Hence, always using the asterisk to mark the limiting case,
we have, from (5), 
\be r^{\ast}_{v=c}(t) >r^{\ast}_{\rm PLC}(t)  \ee

\no for all  $t<0$. Now, from Fig.\,2 we see that, 
for all $ t < 0$,
\be  R(t) >R^{\ast}(t)\;, \hspace{5mm}  
     \dot{R}(t) < \dot{R}^{\ast} (t)\;,    \ee 

\no  
hence, from the definitions (1) and (2), we have, for all
$t<0$, 
\be   r_{v=c}(t) >r^{\ast}_{v=c}(t),\; 
      r_{\rm PLC}(t) <r^{\ast}_{\rm PLC}(t)\;.  \ee

\no Combining the inequalities (8) and (6), we have, for all
$t<0$,
\be   r_{v=c}(t) > r_{\rm PLC}(t) \;. \ee

\no This result is valid for all ($\Lambda_c, \Omega_M>0$) 
models. Combining this with (6), we have that for all
($\Lambda_c, \Omega_M\geq 0$) models, the two curves (1) and
(2) never intersected at any time in the past. Hence we have

\begin{quotation}

\vn {\bf Lemma 2}~~{\em If the universe started expansion
infinitely slowly from either zero or a finite size, then all
the objects we observe now had subluminal recession
velocities at the time of emission }

\end{quotation}

Combining the two Lemmas stated above, I now conclude with
the following statement: ``to the question whether among the
objects we now see, there are some that were receding
faster than light at the time of emission, the answer is
`Yes', if the universe started the expansion with a bang; `No',
if the expansion started infinitely slowly.''

\vn {\bf Acknowledgement}~~I thank Tamara M. Davis for
stimulating correspondence. 

\vn

\bc {\bf References} \ec

\vn Bondi H., 1960, Cosmology. Cambridge University Press. pp.
80, 82 

\vn Davis, Tamara M. et al., 2003, Amer.\ J.\ Phys., 71(4),
358-364

\vn Felten J.E., Isaacman R., 1986, Rev.\ Mod.\ Phys., 58/3,
689-698
 
\vn Kiang T., 1997, Chin.\ Astron.\ Astrophys., 21/1, 1--18. Chinese
version in Acta Astrophys.\ Sinica (Tianti Wuli Xuebao) 1997,
17/3, 225-238

\vn Kiang T., 1987, Quaterly J, R. astron. Soc., 28, 456-471

\end{document}